# Liquid-phase exfoliated indium-selenide flakes and their application in hydrogen evolution reaction

*Elisa Petroni,[#] Emanuele Lago,[#] Sebastiano Bellani, Danil W. Boukhvalov, Antonio Politano, Bekir Gürbulak, Songül Duman, Mirko Prato, Silvia Gentiluomo, Reinier Oropesa-Nuñez, Jaya-Kumar Panda, Peter S. Toth, Antonio Esau Del Rio Castillo, Vittorio Pellegrini, and Francesco Bonaccorso\**

E. Petroni, E. Lago, Dr. S. Bellani, Dr. A. Politano, S. Gentiluomo, Dr. R. Oropeza-Nuñez, Dr. J. K. Panda, Dr. P. S. Toth, Dr. A. E. Del Rio Castillo, Dr. V. Pellegrini, Dr. F. Bonaccorso
Graphene Labs, Istituto Italiano di Tecnologia, via Morego 30, 16163 Genoa, Italy
francesco.bonaccorso@iit.it
E. Petroni, E. Lago, S. Gentiluomo
Dipartimento di Chimica e Chimica Industriale, Università degli Studi di Genova, via Dodecaneso 31, 16146 Genoa, Italy
Dr. M. Prato
Materials Characterization Facility, Istituto Italiano di Tecnologia, via Morego 30, 16163 Genoa, Italy
Prof. D. W. Boukhvalov
Department of Chemistry, Hanyang University, 17 Haengdang-dong, Seongdong-gu, Seoul 04763, South Korea
Dr. Bekir Gürbulak
Department of Physics, Faculty of Sciences, Atatürk University, 25240 Erzurum, Turkey
Prof. Songül Duman
Department of Basic Sciences, Faculty of Sciences, Erzurum Technical University, 25050 Erzurum, Turkey

Keywords: indium selenide, liquid-phase exfoliation, water splitting, hydrogen evolution reaction, electrocatalysis

**Abstract**

**Single- and few-layered InSe flakes are produced by the liquid-phase exfoliation of β-InSe single crystals in 2-propanol, obtaining stable dispersions with a concentration as high as 0.11 g L$^{-1}$. Ultracentrifugation is used to tune the morphology, *i.e.*, the lateral size and thickness of the as-produced InSe flakes. We demonstrate that the obtained InSe flakes have maximum lateral sizes ranging from 30 nm to a few μm, and thicknesses ranging from 1 to 20 nm, with a max population centred at ~ 5 nm, corresponding to 4 Se-In-In-Se quaternary layers. We also show that no formation of further InSe-based compounds (such as In$_2$Se$_3$) or oxides occurs during the exfoliation process. The potential of these exfoliated-InSe few-layer flakes as a catalyst for**

**hydrogen evolution reaction (HER) is tested in hybrid single-walled carbon nanotubes/InSe heterostructures. We highlight the dependence of the InSe flakes morphologies, *i.e*., surface area and thickness, on the HER performances achieving best efficiencies with small flakes offering predominant edge effects. Our theoretical model unveils the origin of the catalytic efficiency of InSe flakes, and correlates the catalytic activity to the Se vacancies at the edge of the flakes.**

# 1. Introduction

As a result of the energy crisis and environmental pollution, one of the main challenges for today's society is to produce highly efficient and low-cost renewable energy sources[1]. Hydrogen is one of the most promising clean energy carriers, and it has consequently attracted an increasing amount of interest concerning both energy production and storage.[2–4] Specifically, electrochemical water splitting is viewed as an efficient and scalable method for viable $H_2$ production.[5,6] To date, Pt-based systems are the most efficient electrocatalysts for a hydrogen evolution reaction (HER) since they have an overpotential at a cathodic current density of 10 mA cm$^{-2}$ ($\eta_{10}$) at nearly 50 mV.[7–9] Nevertheless, the high cost (> 30 USD g$^{-1}$)[10] and the limited availability (< 0.005 ppm)[11] of Pt make the quest to find alternative cheaper materials inevitable.[8,9,12–14]

In this regard, layered materials such as transition-metal dichalcogenides (TMDs),[15–19] transition-metal carbides[20] and oxides[21–24] are promising Earth-abundant candidates as HER-electrocatalysts. Furthermore, the exfoliation of bulk layered materials in atomically thick flakes is effective for improving the HER efficiency compared to the bulk counterpart due to the increase in the number of exposed active sites, *i.e.*, the defects or vacancies that are mainly located at the edges of the 2D flakes.[8,9,25–27]. According to theoretical predictions,[28–31] the family of III−VI layered compounds, MX (M = Ga, In; X = S, Se, Te), are good candidates for hydrogen production.[32] MX compounds are wide bandgap semiconductors, with band gaps ranging from 1.2 to ~ 3.0 eV at room temperature (RT),[32–35] made of stacked quaternary layers of X–M–M–X atoms that are held together by van der Waals interactions.[32,36] Similarly to exfoliated TMDs, such as the most studied $MoS_2$[15–19] and $MoSe_2$,[15–19] the unsaturated chalcogen-edges in the natural phase of MX compounds can be HER electrocatalytically active, with Gibbs free energy ($\Delta G$) of adsorbed protons ($H_{ads}$) close to zero.[15–19] Notably, density functional theory (DFT) calculation evidenced that MX monolayers exhibit band edges located at energetically favourable positions for solar water splitting,[28] *i.e.*, conduction band minimum energy higher than the standard reduction potential of $H^+/H_2$, which is –4.44 eV vs. vacuum energy level at pH = 0, and a valence band maximum energy lower than the standard oxidation potential of $O_2/H_2O$, which is –5.67 eV vs. vacuum energy level at pH = 0.[28] Among these monolayers, indium selenide (InSe) has a conduction band edge energy of –4.14 eV vs. vacuum energy level,[28] which is closer to the HER standard reduction potential than that of other members of the III-VI family.[28] Based on this consideration, the investigation of the HER–electrocatalytic activity of InSe, as well as that on the other exfoliated MX compounds, is of utmost interest. Up to now, early experiments carried out in acid electrolyte have been reported for gallium sulfide (GaS) flakes obtaining $\eta_{10}$ ~ 570 mV.[37]

Mechanically exfoliated flakes of InSe, as well as its heterostructures,[38] have been widely used for devising optoelectronic devices and photodetectors displaying high on/off ratios (~ $10^8$),[39] broad-band response (from UV-Vis to NIR)[40] and ultra-fast response time (up to 60 μs by avalanche effect).[41] Nevertheless, electrocatalytic applications are still missing.

With regard to InSe, these flakes have been widely used for designing electronic and optoelectronic devices.[38–41] In particular, transistors and photodetectors made of few-layered InSe display high ON/OFF ratios (~ $10^8$),[39] broad-band response (from UV-Vis to NIR)[40] and ultra-fast response time (up to 60 μs by avalanche effect).[41] Vertical graphene/InSe heterostructures have also been devised, reaching photoresponsivity values of $10^5$ A $W^{-1}$.[38] Current InSe-based electronic and optoelectronic devices have been produced with mechanically exfoliated crystals.[38–41] However, micromechanical cleavage is not an industrially-relevant approach, especially for wide-scale technology transfer, since it suffers from a low-throughput.[42,43] Wafer-scale production methods have been recently implemented for the deposition of 2D metal oxide (*i.e.* $Ga_2O_3$, $Al_2O_3$)[44] and MX compounds (*i.e.* GaS),[45] These methods exploit the formation of an oxide thin surface layer in metal-based liquid through a self-limiting reaction naturally occurring in ambient conditions.[44,45]

Despite the latter progress, exfoliated InSe has not been exploited yet in electrocatalytic applications.[28] In this context, wafer-scaled 2D materials are not required,[15-19] and liquid-phase exfoliation (LPE) is a scalable and cheap technique with respect to other synthesis approaches,[42,43] enabling the exfoliation of layered bulk materials in a liquid medium (*e.g.*, solvents[42,46–48] and surfactant or polymer solutions[42,46–48]) by applying an external driving force such as ultra-sonication.[42,46–48] Thus, LPE avoids the need for growing substrates, long processing time and synthesis precursors/ligands, which are typical constraints of the chemical vapour deposition (CVD) processes and colloidal synthesis.[42,49,50] It is worth noting that the LPE-produced dispersion is heterogeneous both in terms of the lateral size (meant as the maximum lateral size of the single flake) and thickness of the as-produced flakes.[43] Thus, a purification process is needed to obtain a stable dispersion containing a high percentage of single-layered flakes.[43,51] This requires an ultra-centrifugation of the LPE-produced dispersion.[52–55] This process, commonly known as 'sedimentation based-separation' (SBS), separates flakes based on their sedimentation rate in response to a centrifugal force acting upon them.[56] By tuning the centrifugal forces, it is possible to obtain dispersions with flakes that have different lateral sizes and thicknesses.[42,43] For these reasons, LPE followed by SBS is an ideal up-scalable production technique for making 2D materials that are competitive in terms of energy storage[42,43] and conversion,[42,43] as well as in other fields of application.[42,43,57]

The most common solvent used to exfoliate layered crystals is N-methyl-2-pyrrolidone (NMP)[47,57,58] because of its suitable surface energy and Hansen's solubility parameters (HSPs).[58] Nevertheless, NMP is a teratogenic solvent (Health code ≥ 2, NFPA704)[59] with a high boiling point (202 °C)[60]. These two characteristics raise concerns for its applicability for the large-scale production of 2D crystals for technological applications.[58] In particular, the boiling point of the solvent has to be taken into account with regard to the realization of high-performance devices.[52,61,62] The removal of the solvent is crucial, and this is achieved by heating up the sample or the entire device above its boiling point, but this process could either degrade the sample or damage the device.[52,58,61,62] Notably, when heated, NMP leaves contaminants as it degrades,[58,63,64] which subsequently reduces the device performance.[52,65] For these reasons, non-toxic solvents that are easily handled (*i.e.* solvents that have a low boiling point)[58] and that have a low environmental impact[61,62] are particularly attractive. This issue, however, has not been addressed yet with regard to InSe. In fact, a recent study reports the LPE of β-InSe from bulk powder into few-layered flakes,[57] which have been used to develop a photodetector, by using NMP as a solvent.

Here, we demonstrate the LPE of β-InSe in 2-propanol (IPA), a non-toxic solvent (Health code 1; NFPA704)[59] with a low boiling point (82.6 °C),[60] as an effective route for the exfoliation of β-InSe crystals in atomically thick flakes, without the formation of other chemical species, *i.e.* $In_2Se_3$ or $In_2O_3$. We propose the use of LPE InSe flakes as electrocatalysts for HER displaying pH-universal HER activity, which is dependent on the flakes morphology, *i.e.*, surface area and thickness. In particular InSe flakes of smaller surface area and thickness have the lowest reported values of $\eta_{10}$ among the MX compounds ($\eta_{10}$ of 549 mV at pH = 1 and 451 mV at pH=14).[37,66] Finally, to unravel the source of the catalytic efficiency of InSe flakes, a theoretical model based on DFT simulations has been developed, which correlates the catalytic activity of InSe flakes with chalcogen edge-vacancies.

## 2. Results and discussion

### 2.1. Material production and characterization

β-InSe single-crystals, produced by the modified Bridgman-Stockbarger method[67] (Figure 1a), are exfoliated by means of LPE in IPA. The use of this solvent has significant advantages in material processing and analysis due to its nontoxicity and low boiling point.[60] Moreover, IPA has been successfully used to exfoliate layered compounds such as GaS,[37,68] GaSe,[68] and TMDs.[47] Sonication (Figure 1c) and ultra-centrifugation (Figure 1d) of β-InSe single-crystals are performed to obtain a stable pale orange dispersion of exfoliated β-InSe flakes (Figure 1e and bottom inset of Figure 1f). With the aim to investigate the HER activity dependence on the β-InSe flakes' morphology, we prepared three different samples which were selected by varying the ultra-centrifugation speeds, *i.e.* 200*g* (ex-InSe 200*g*), 1000*g* (ex-InSe 1000*g*) and 4000*g* (ex-InSe 4000*g*). In the following text, structural and morphological analyses refer to the ex-InSe 1000*g* sample, simply named 'ex-InSe', which was expected to be a trade-off in the physicochemical properties between the slower and the higher ultra-centrifugation speeds. The morphological and optical characterization of ex-InSe 200*g* and ex-InSe 4000*g* are reported in the Supporting Information.

In order to evaluate the concentration of the dispersed ex-InSe flakes, optical extinction measurements were performed. Figure 1f shows a typical ex-InSe extinction spectrum which is characterized by two features at ~ 275 nm and at ~ 360 nm that are related to the maxima in the imaginary part of the dielectric function that is connected to the direct transitions from the valence band to the conduction band.[69] The corresponding wavelength of both peaks is shifted toward lower values by decreasing the average flake thickness (Figure S1) according to the planar quantum confinement of photo-excited carriers.[35]

The concentration of the dispersion is estimated by the Beer-Lambert law:

$$E = \varepsilon c l \quad (1)$$

in which $E$ is the extinction, $\varepsilon$ is the optical extinction coefficient, $c$ is the concentration and $l$ is the path length. Optical extinction measurements, at known concentrations of ex-InSe dispersed flakes in IPA, allow the optical extinction coefficient to be estimated. The slope of this curve provides $\varepsilon$ = 580 L g$^{-1}$ m$^{-1}$ at 600 nm (top inset to Figure 1f). At this wavelength, the extinction coefficient is found to be size-independent, as optical extinction spectra of samples sorted by different ultra-centrifugation speeds overlap above 550 nm (Figure S2). By using the experimentally derived $\varepsilon$ value, the concentration of ex-InSe sample, properly diluted (1:10), is 0.11 g L$^{-1}$. This result is further confirmed by weight measurement of the solid material content in a known dispersion volume of ex-InSe (measured concentration value of 0.13 ± 0.02 g L$^1$).

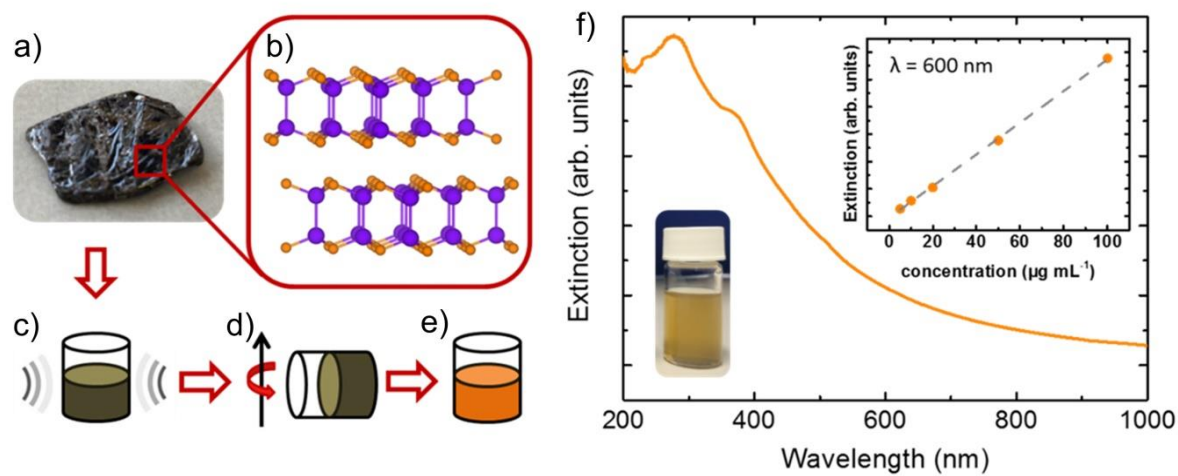

**Figure 1.** Production of β-InSe ultrathin flakes by LPE. a β-InSe single crystal, grown by the modified Bridgman-Stockbarger method; b atomic structure of β-InSe; c, d and e schematics of the LPE process: sonication; ultra-centrifugation; and stable dispersion of exfoliated β-InSe flakes, respectively. f extinction spectrum of the ex-InSe sample. Top inset: optical extinction at 600 nm vs concentration for ex-InSe. Bottom inset: photograph of ex-InSe dispersion in IPA.

The effective exfoliation of β-InSe in IPA is evaluated by means of morphological, chemical and structural analyses (Figure 2) in order to confirm the production of few-layers β-InSe flakes together with the preserved elemental composition and lattice structure. Representative images of transmission electron microscopy (TEM) and atomic force microscopy (AFM) for ex-InSe are reported in Figures 2a and 2b, respectively. Statistical analysis, fitted by a log-normal distribution[70] (Figure 2c), indicates that ex-InSe have a typical lateral size of ~ 113 nm (with a log-normal standard deviation of 0.84), a corresponding average surface area of ~ 6.1·10$^{-3}$ μm$^2$ (see the surface area statistical analysis in the inset to Figure 2c), and a most probable thickness of 4 nm (with a log-normal standard deviation of 0.54). Thus, according to a β-InSe monolayer step thickness of about 0.9 nm,[50,71] single/few-layered β-InSe flakes are effectively produced, resulting in a distribution peak at ~ 4 quaternary Se–In–In–Se layers. In the inset of Figure 2b an AFM image of single-layer ex-InSe flakes is also reported. In fact, its thickness of ~ 1.5 nm experimentally agrees with that of an InSe monolayer (expected bilayer thickness is ~ 1.8 nm).[50,71] The morphological characterization of ex-InSe 200$g$ and ex-InSe 4000$g$ are reported in Figure S3. As expected, the slower (higher) ultra-centrifugation speed has a larger (smaller) overall flake size than the ex-InSe 1000$g$ sample, and has an average lateral size of ~ 160 nm (75 nm), surface area of ~ 11.2·10$^{-3}$ μm$^2$ (2.6·10$^{-3}$ μm$^2$) and thickness of ~ 6.85 nm (2.61 nm). Figure 2d shows X-ray photoelectron spectroscopy (XPS) spectra of crystalline β-InSe dispersed in IPA (c-InSe) and ex-InSe for both the In and Se 3d core levels. Both the c-InSe and ex-InSe spectra show two peaks in the In 3d spectral range, which relate to the spin-orbit splitting of the d orbitals (3d$_{5/2}$ at 444.2 eV and 3d$_{3/2}$ at 451.8 eV),[72] and four peaks in the Se 3d spectral range, which are associated to the spin-orbit splitting of Se$^{2-}$ and Se$^0$ (3d$_{5/2}$ at 53.7/54.5 eV and 3d$_{3/2}$ at 54.6/55.3 eV).[72] The high binding energy components of Se are ascribed to surface modifications due to exposure to ambient conditions.[50,73] Elemental analysis of the XPS spectra supports the preservation of the compound stoichiometry after the LPE, as also confirmed by the TEM energy dispersive X-ray spectroscopy (STEM-EDS, Figure S4) that was performed on isolated flakes. Nevertheless, the exfoliation process in IPA causes a slight oxidation of selenium (9.3 %), as is indicated by the features located at 58.6 and 59.5 eV, which are ascribed to the 3d$_{5/2}$ and 3d$_{3/2}$ core levels, respectively, in the spin-orbit splitting of Se$^{4+}$.[72] In addition, the presence of Se vacancies in exfoliated flakes is evaluated by a comparison of In and Se core levels in c-InSe and ex-InSe samples. On the basis of the quantitative XPS analysis, we can support that an increase of ~ 25% in Se vacancies is introduced upon exfoliation with respect to the bulk counterpart.

The acquired X-ray diffraction (XRD) patterns (Figure 2e) demonstrate, for both the bulk (cyan trace) and the exfoliated samples (orange trace), the presence of a β-InSe hexagonal structure ($D^4_{6h}$ symmetry, P63/mmc space group) with calculated lattice parameters of $a = b = 4.005 \pm 0.004$ Å and $c = 16.660 \pm 0.004$ Å. These findings are in agreement with the International Centre of Diffraction Data [ICDD 98-018-5172]. In particular, in the case of c-InSe, only peaks belonging to the (00$l$) plane family were observed, indicating that there are highly oriented crystalline flakes along the $c$-axis. On the other hand, for the XRD pattern of the ex-InSe sample, planes with different orientations are visible (*e.g.* 010, 011 and 110). Raman spectroscopy measurements were also carried out with the aim of unambiguously demonstrating the absence of other chemical species, *e.g.* $In_2Se_3$ or $In_2O_3$,[74,75] during the exfoliation process or during exposure to ambient gases. Accordingly to the aforementioned β-InSe unit cell structure, counting three degrees of freedom for each atom, β-InSe has 24 vibrational modes at the centre Γ of the Brillouin zone.[76] Of these 24 modes, six are Raman active ($2A_{1g}$, $2E_{1g}$, $E^1_{2g}$, $A_{2u}$)[75,76] and their corresponding atomic vibrations are shown as an inset to **Figure 2f**. As was evidenced by measuring the Raman spectra at RT (**Figure 2f**), the six first-order modes are visible in both the c-InSe and ex-InSe samples. They are classified as $E^1_{1g}$ at 40 cm$^{-1}$, $A^1_{1g}$ at 117 cm$^{-1}$, $E^2_{1g}$ at 178 cm$^{-1}$, $E^1_{2g}$ at 200 cm$^{-1}$, $A_{2u}$ at 210 cm$^{-1}$, $A^2_{1g}$ at 227 cm$^{-1}$, which is in agreement with previous theoretical[77] and experimental works.[75] Second-order modes are also visible at RT and result from overtones of $E^1_{2g}$ (402 cm$^{-1}$) and $A_{2u}$ (423 cm$^{-1}$).[78] It is worth noting that the peak positions of all modes have no significant energy shift in the ex-InSe sample as is the case for c-InSe (Figure S5), which is in agreement with previous experimental studies.[75,79] Interestingly, no modes relating to other InSe-based compounds are observed, confirming that the stoichiometry is preserved without the formation of other chemical species during the exfoliation process.[74,75] These conclusions are further supported by XRD patterns, displayed in **Figure 2e**. The morphological and optical characterizations, *e.g.*, optical absorption, lateral size and thickness analyses, of ex-InSe 200$g$ and ex-InSe 4000$g$ are reported in the Supporting Information.

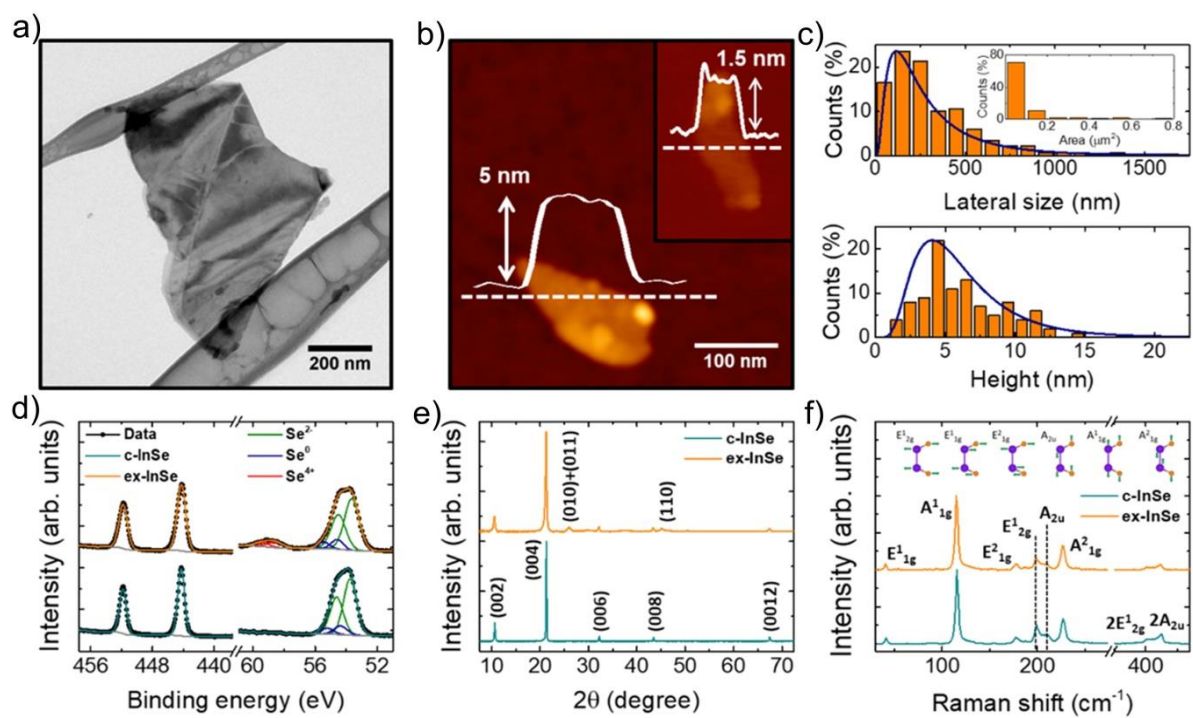

**Figure 2.** Morphological, chemical and structural characterization of ex-InSe flakes. a Representative TEM image of an isolated ex-InSe flake; b Representative AFM image of an isolated ex-InSe flake. Height profile (solid white lines) of the indicated section (white dashed lines) is also shown. Inset: AFM image of an isolated ex-InSe flake with thickness of ~ 1.5 nm; c Lateral size and thickness statistical analyses for ex-InSe flake dispersion. Inset: statistical analysis for ex-InSe surface area; d In-3d and Se-3d core-level spectra for c-InSe (cyan trace) and ex-InSe (orange trace) samples; e XRD spectra for c-InSe (cyan trace) and ex-InSe (orange trace) samples; f Raman spectra for c-InSe (cyan trace) and ex-InSe (orange trace) samples. Inset: schematic of atomic vibrations for β-InSe Raman active modes.

## 2.2. Electrocatalyst fabrication and characterization

To evaluate the HER activity of β-InSe, both bulk and flake dispersions, purified at different ultra-centrifugation speeds (*i.e.* 200$g$, 1000$g$ and 4000$g$), were tested. The β-InSe samples were deposited on single wall carbon nanotube (SWCNTs)-based films, which are used as flexible current collectors. As demonstrated by recent studies on carbon-based hybrid heterostructures,[80–85] SWCNTs-based films can increase the electron accessibility to the HER-active sites, speeding up the HER-kinetics.[86–88] compared to glassy carbon (GC) rigid electrodes. In our studies, we observed that the low adhesion of deposited β-InSe atop GC electrodes does not allow for a proper evaluation of the electrocatalytic properties of the material, *i.e.* the deposited β-InSe becomes detached from the substrate during the measurements, especially in the exfoliated samples. Additionally, the use of SWCNT-based buckypaper provides stability for the electrocatalysts in aqueous electrolytes. This permits to avoid the use of ion conducting catalyst binders (typically Nafion®[89] in acid solution and Tokuyama AS-4®[90] in the alkaline one), which can be expensive[91,92] or detrimental to the electrocatalytic activity of the catalyst.[93,94] The hybrid heterostructures formed by commercial SWCNTs (mass loading 1.2 mg cm$^{-2}$) and β-InSe samples (mass loading 1.2 mg cm$^{-2}$ for each sample) were produced through a sequential vacuum filtration deposition on nylon membranes with flexible binder-free active electrodes, which are compatible with high-throughput scalable industrial manufacturing. **Figures 3a** and **3b** display high-resolution scanning electron microscopy (HR-SEM) cross-sectional images of the as-produced c-InSe and ex-InSe based heterostructures (SWCNT/c-InSe and SWCNT/ex-InSe), respectively. Both the SWCNT/c-InSe and the SWCNT/ex-InSe electrocatalysts show a bilayered-like architecture with a distinct separation between the SWCNT-based collector and the β-InSe-based active film (~ 3 μm film thickness for ex-InSe sample). Thus, as is also evidenced by top-view images (insets to **Figures 3a** and **3b**), no penetration of β-InSe within the SWCNT-based buckypaper was observed, suggesting that the role of SWCNT film is that of a bare current collector (Figure S6 for SWCNT electrochemical characterization). The HER-activities of the as-produced InSe-based electrocatalysts were evaluated in both acid (0.5 M $H_2SO_4$, pH = 1) and alkaline (1 M KOH, pH = 14) electrolytes (**Figure 3**). The HER under acidic conditions consists of two steps: $H_{ads}$ at the catalytic sites of the electrode (Volmer reaction, $H_3O^+ + e^- \rightleftarrows H_{ads} + H_2O$), and either an electrochemical desorption step (Heyrovsky reaction, $H_{ads} + H_3O^+ + e^- \rightleftarrows H_2 + H_2O$) or a recombination step (Tafel reaction, $2H_{ads} \rightleftarrows H_2$).[95,96] The occurrence of the two reactions depends on the Gibbs free energy of the process; the one which has the energy value closer to 0 eV is promoted.[95,96] In alkaline media, the $H_{ads}$ is formed by the discharge of $H_2O$ (Volmer reaction, $H_2O + e^- \rightleftarrows H_{ads} + OH^-$).[95,96] Then, either the Heyrovsky step ($H_2O + H_{ads} + e^- \rightleftarrows H_2 + OH^-$) or the chemical Tafel step ($2H_{ads} \rightleftarrows H_2$) occurs, depending on the Gibbs free energy value of the two steps.[95,96] The overpotential at a 10 mA cm$^{-2}$ cathodic current density ($\eta_{10}$), the Tafel slope and the exchange current density ($j_0$) are typical figures of merit (FoM) to evaluate the HER performance of electrocatalysts.[95,96] In detail, the Tafel slope and $j_0$ are estimated from the linear portion of the Tafel plots (overpotential vs. current density curves) according to the Tafel equation (see the Experimental Section for further details).[95,96] In particular, the Tafel slope is used to evaluate the HER mechanism at the electrode/electrolyte interface, while $j_0$ relates to the amount of the available active sites for HER.[95,96] In fact, for insufficient $H_{ads}$ surface coverage, the Volmer reaction is the rate-limiting step of HER, and a theoretical Tafel slope of 120 mV dec$^{-1}$ is expected.[95,96] Conversely, in the limit of high $H_{ads}$ surface coverage, the HER-kinetics is dominated by Heyrovsky or Tafel reactions, and has Tafel slopes of 40 or 30 mV dec$^{-1}$, respectively.[95,96]

To relate the morphology of β-InSe flakes to HER performance, both c-InSe and samples ultra-centrifuged at different speeds, *i.e.*, ex-InSe 200*g*, ex-InSe 1000*g* and ex-InSe 4000*g*, were tested. The current density *vs.* potential relative to reversible hydrogen electrode (RHE) curves of β-InSe based electrocatalysts under both acidic and alkaline conditions are reported in **Figures 3c** and **3d**, respectively. As it has been previously observed on layered crystals,[97,98] including GaS,[37] the HER-activity of β-InSe flakes increases when the flakes' size (lateral size and thickness) decreases. Indeed, the $\eta_{10}$ value scales from 581 mV (540 mV) for c-InSe, to 561 mV (483 mV) for ex-InSe 200*g*, to 556 mV (471 mV) for ex-InSe 1000$g$, to 549 mV (451 mV) for ex-InSe 4000$g$ in an acid (alkaline) solution. As it has been theoretically [99,100] and experimentally[37,97,98,101] demonstrated, this effect is mostly due to the increased number of HER-active sites. In fact, the reduction in the lateral size and thickness of the flake, which occurs when the ultra-centrifugation speed is enhanced, exposes the liquid electrolyte to a larger amount of active edge defects and vacancies than those of the bulk counterpart.[99–101] A rigorous kinetic analysis of the HER, *i.e.* the establishment of the Tafel slope and $j_0$, is not conducted here because of the unambiguous results that were obtained in the presence of high-surface area SWCNT buckypaper, which gives a capacitive current density even for a low voltage scan rate ($\leq$ 5 mV s$^1$).[102] Nevertheless, it is worth noting that the calculated $j_0$ value (Table ST1) for the exfoliated β-InSe samples increases from 0.18 μA cm$^2$ (0.45 μA cm$^2$) for ex-InSe 200$g$ to 0.42 μA cm$^2$ (5.8 μA cm$^2$) for ex-InSe 4000$g$ in an acid (alkaline) solution, suggesting that the catalytic activity is dominated by the edge sites. This statement is further confirmed by Electrochemical Impedance Spectroscopy (EIS) analysis, which correlates the increase double-layer capacitance ($C_{dl}$), *i.e.* the specific surface area, of the electrodes with the increment of the ultra-centrifugation speed (**Figure S17**).[103] Experimental Tafel slopes, reported together with their corresponding Tafel plots for which pH = 1 (**Figure 3e**) and pH = 14 (**Figure 3f**), have values higher than 120 mV dec$^{-1}$ for all the β-InSe samples in both an acid and alkaline solution, indicating that the Volmer reaction step determines the rate of the HER. Moreover, Tafel slope values exceeding the Volmer slope of 120 mV dec$^1$ have been previously described in theoretical studies as not having ideal behaviour due to the anion adsorption or non-uniform distribution of the surface electric field on rugged electrodes.[104–106] Previous electrochemical characterizations of MX compounds have been performed on GaS[37] and GaSe.[66] The latter, tested in its bulk form, has a poor electrocatalytic activity; $\eta_{10}$ > 1V on GC in an acid solution.[66] Conversely, exfoliated GaS deposited on pyrolytic carbon reaches a Tafel slope value of 85 mV dec$^{-1}$ and an overpotential $\eta_{10}$ ~ 570 mV under acidic conditions.[37] Thus, LPE-produced atomically thick layers of InSe exhibit the lowest $\eta_{10}$ value among the MX family.

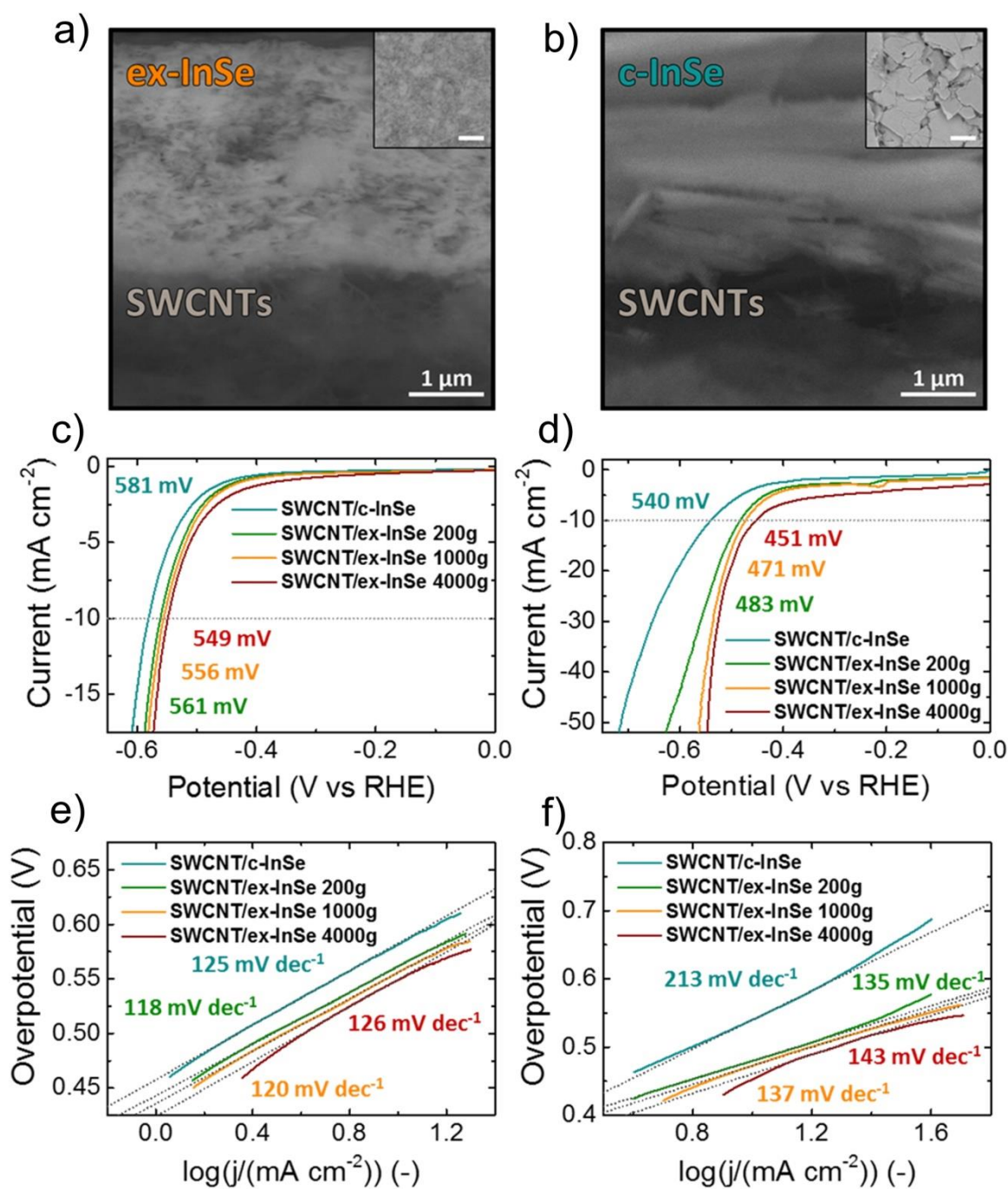

**Figure 3.** Morphological and electrochemical characterization of SWCNT/β-InSe hybrid electrocatalysts. a Cross-sectional HR-SEM image of SWCNT/c-InSe heterostructure (top view in inset, scale bar is 1 μm); b Cross-sectional HR-SEM image of SWCNT/ex-InSe 1000*g* heterostructure (top view in inset, scale bar is 1 μm); c Current density vs. potential

relative to RHE curves for SWCNT/c-InSe, SWCNT/ex-InSe 200*g*, SWCNT/ex-InSe 1000*g* and SWCNT/ex-InSe 4000*g* samples at pH = 1; d Current density *vs.* potential relative to RHE curves for SWCNT/c-InSe, SWCNT/ex-InSe 200*g*, SWCNT/ex-InSe 1000*g* and SWCNT/ex-InSe 4000*g* samples at pH = 14; e Tafel plots of the same β-InSe based electrocatalysts shown in panel c. Linear fits (dashed lines) and corresponding Tafel slope values are reported; f Tafel plots of the same β-InSe based electrocatalysts shown in panel d. Linear fits (dashed lines) and corresponding Tafel slope values are reported.

In order to unveil the origin of the catalytic activity of InSe, DFT calculations were carried out based on a theoretical model using a 64-atom supercell ($In_{32}Se_{32}$). The energy of the reactions that rule the adsorption of water fragments (*i.e.* $H^+$ and $OH^-$), as well as the energy barriers of the $H_2O$ dissociation (*i.e.* $H_2O + e^- \rightarrow H_{ads} + OH^-$) on InSe monolayer were calculated (**Table 1**). Notably, the latter reaction coincides with the Volmer step in alkaline environment. This step could also be concomitant with the adsorption of $OH^-$, whose desorption is effective in promoting HER.[107] To estimate the differential enthalpy (ΔH) that is associated to the Volmer step, the total energy of the InSe monolayer with two physisorbed water molecules (**Figure 4a**) is compared to the total energy of the same system with chemisorbed H, *e.g.* $H_{ads}$ (**Figure 4b**). The presence of a second water molecule represents a minimalistic model of the water environment. Moreover, in order to have a complete representation of the HER process in an alkaline environment, the differential enthalpy associated to the adsorption of $OH^-$ ($OH_{ads}$) is evaluated (**Figure 4c**). The differential Gibbs free energy is evaluated using the formula:

$$\Delta G = \Delta H - T\Delta S + ZPE - eU, \qquad (2)$$

In which T is the RT and ΔS is the changes of entropy of the system, ZPE is the zero point energy and U is the overpotential (0.83 V for reaction under alkaline conditions and 1.23 V for reaction under acidic conditions).[108–110]

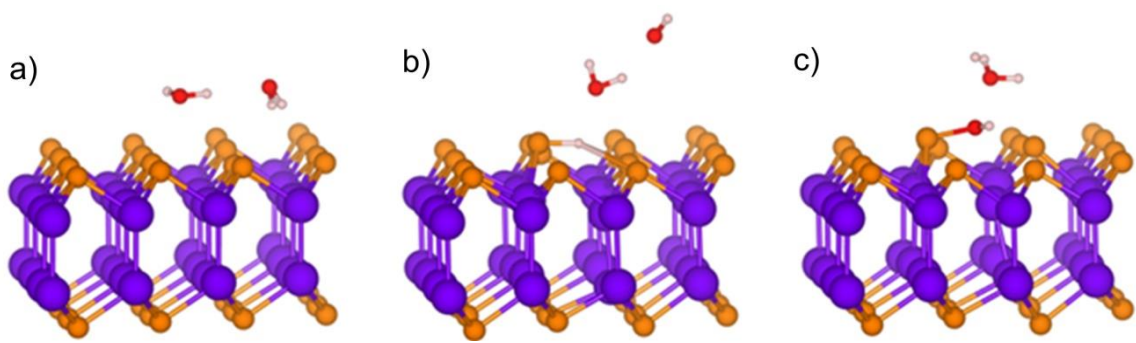

**Figure 4.** Scheme of the DFT models used for investigating InSe catalytic activity. **a** InSe supercell with a physisorbed pair of water molecules; **b** and **c** chemisorption on InSe monolayer of $H_{ads}$ and $OH_{ads}$, respectively.

The results of the DFT calculations (**Table 1**) for pristine InSe monolayers indicate that the energy required for an $H^+$ adsorption is unfavourable ($\Delta G > 2$ eV) in both acidic and alkaline environments. Nevertheless, the presence of thermodynamically unavoidable lattice defects, such as Stone-Wales defects or Se vacancies, makes the adsorption of water fragments exothermic. In particular, the HER activity can be mainly addressed to the Se vacancies, which are naturally present in β-InSe single crystals.[36,67,111] In fact, taking into account the Se vacancies, the $\Delta G$ for $H_{ads}$ in both acidic and alkaline environments is closer than those of the un-defected and Stone-Wales defected InSe monolayer; they have the ideal value of $\Delta G = 0$ eV, which is an energetically feasible HER condition.[112,113] A differential Gibbs free energy close to 0 eV indicates that the $H_{ads}$ free energy is close to that of the reactant or product. Low $\Delta G$ suggests that there is a strong bonding of $H_{ads}$, while a high $\Delta G$ implies that $H_{ads}$ are weakly bonded to the catalyst surface. Both conditions negatively affect the HER kinetics.[113] In addition, the $\Delta G$ for $OH^-$ adsorption is reduced in presence of Se vacancies (–0.84 eV) with respect to the pristine case (+2.47 eV). Thus, the beneficial $OH^-$ desorption in an alkaline environment is favoured by the defected InSe monolayer. As aforementioned, the amount of Se vacancies has been estimated by a comparative evaluation of β-InSe core levels by XPS in c-InSe and ex-InSe samples, reporting an increase of 25% in the exfoliated flakes.

Finally, the reduction of ΔG for $H_{ads}$ in presence of Se vacancies is compatible with the $\eta_{10}$ downward trend that was observed in both the acid and alkaline environments when the flakes' lateral size is reduced. As a result, the edge-related HER activity is therefore linked to an increase in the number of exposed Se vacancies, which are mostly placed at the edge of the exfoliated flakes.

**Table I.** Calculated values of the differential enthalpy (ΔH) and Gibbs free energy (ΔG) for $H_{ads}$ and $OH_{ads}$ in both acidic and alkaline environments.

|  | ΔH (eV) | | ΔG (eV) | | |
|---|---|---|---|---|---|
| Condition | $H_{ads}$ | $OH_{ads}$ | $H_{ads}$ acid | $H_{ads}$ alkali | $OH_{ads}$ alkali |
| InSe-monolayer | +3.34 | +2.95 | +2.23 | +2.63 | +2.47 |
| Stone-Wales defect | -1.05 | -0.41 | -2.16 | -1.76 | -0.89 |
| Se-vacancy | +0.65 | -0.36 | -0.46 | -0.06 | -0.84 |

## 3. Conclusion

In conclusion, effective exfoliation of InSe single crystals by means of LPE in a non-toxic solvent has been demonstrated, obtaining a stable dispersion of atomically thick InSe flakes with an IPA concentration up to 0.11 g $L^{-1}$. Transmission electron and atomic force microscopy analyses allowed the exfoliated flakes' morphology to be evaluated; they had an average thickness of ~ 5 nm, corresponding to ~ 4 InSe layers, a maximum lateral size of ~ 113 nm and an average surface area of ~ $6.1 \cdot 10^{-3}$ μm². Raman and XRD analyses confirmed the preservation of the crystal structural and chemical properties in the exfoliated flakes, excluding the formation of spurious compounds (*i.e.* $In_2Se_3$ or $In_2O_3$). In particular, XRD patterns evidence that typical hexagonal lattice structure of β-InSe is preserved after the exfoliation, excluding the presence of polytypes with different symmetry (*i.e.* γ-InSe, rhombohedral lattice structure). Moreover, an evaluation of Se vacancies amount in the exfoliated flakes has been performed by means XPS measurements. These revealed an increase of vacancies of ~ 25% with respect to that of the bulk counterpart. Together with this result, XPS analysis has shown a mild oxidation (< 10%) of the exfoliated sample, thus confirming the preservation of the β-InSe chemical composition during the exfoliation process. Thus, LPE in IPA has been demonstrated to produce InSe flakes with thickness at the atomic scale, confirming its crucial role in the application and technological fields as a competitive and environmentally friendly production technique. In particular, electrochemical $H_2$ production was evaluated here as a possible application for InSe flakes. Hybrid and flexible heterostructures composed by SWCNTs and InSe flakes with different morphologies (*i.e.*, different lateral size and thickness) were tested under both acidic and alkaline conditions, and the smallest InSe flakes obtained a promising overpotential $\eta_{10}$ of 549 mV and 471 mV, respectively. These results pave the way for the use of InSe flakes in water splitting and for further optimization as flexible and cost-effective (photo-)electrocatalysts. Moreover, a clear size effect is observed in the InSe HER-performance, suggesting that the active sites are located at the edges of the InSe flakes. This observation is related to the presence of Se vacancies at the edges of the flakes and is demonstrated by means of DFT calculations. In fact, theoretical results indicated a reduction in the exothermic ΔG in the defected InSe monolayer in comparison with its pristine counterpart, reaching ΔG values of -0.46 eV and -0.06 eV for acid and alkaline ambient, respectively. To summarize, the exfoliated InSe flakes have been demonstrated to be suitable materials for catalysis, which extends their application scenery, mostly based, currently, on electronic and optoelectronic devices.[32]

## 4. Experimental Section

*Exfoliation of bulk β-InSe*

β-InSe single crystals were grown by the modified Bridgman-Stockbarger method (details on the growth in Supporting Information).[67] Exfoliated β-InSe flakes were produced through LPE[47], followed by SBS,[42] in IPA (ACS Reagent, ≥ 99.8%, Sigma Aldrich). In short, 40 mg of β-InSe single crystals are pulverized in a mortar and, once added to 20 mL of IPA, ultra-sonicated in a sonicator bath (Branson® 5800 cleaner, Branson Ultrasonics) for 6 h, keeping the bath temperature in the range of 25 °C – 35 °C. The resulting dispersion was ultra-centrifuged at 200$g$ (1000 rpm), 1000$g$ (2500 rpm) or 4000$g$ (5000 rpm) (in a Beckman Coulter Optima™ XE-90 with a SW32Ti rotor) for 30 min at 15 °C in order to separate un-exfoliated and thick β-InSe flakes (collected as sediment) from the as-prepared dispersion. Then, 80% of the supernatant, containing small and thin β-InSe flakes, was collected by pipetting.

*Material characterization*

Optical extinction spectroscopy was carried out by a Cary Varian 5000UV-Vis. In order to measure the extinction spectra, exfoliated β-InSe flake dispersions in IPA were diluted 1:10 with the pure solvent. For each sample, the extinction spectra (absorbed plus scattered light) of the pure IPA were subtracted from the sample spectrum. The optical extinction coefficient was determined by using the Beer-Lambert law ($E = \varepsilon c l$, in which $E$ is the optical extinction at 600 nm, $\varepsilon$ is the extinction coefficient, $c$ is the concentration of the exfoliated β-InSe flakes and $l$ is the path length of the quartz cuvette, 0.01 m).

TEM and Scanning TEM (STEM) imaging and energy-dispersive X-ray spectrometry (EDS) elemental mapping were carried out on a JEOL JEM 1400Plus microscope, operating at 120 kV, equipped with a LaB$_6$ thermionic source, a Gatan CCD camera Orius 830 and a JEOL Dry SD30GV silicon-drift detector (SDD). Samples for the TEM measurements were prepared by drop-casting the exfoliated β-InSe flake dispersions onto carbon-coated Cu grids with successive drying under vacuum overnight.

Atomic force microscopy images were acquired using a Nanowizard III (JPK Instruments, Germany) mounted on an Axio Observer D1 (Carl Zeiss, Germany) inverted optical microscope. The AFM measurements were carried out using PPP-NCHR cantilevers (Nanosensors, USA) with a nominal tip diameter of 10 nm. A drive frequency of ~ 295 kHz is used. Intermittent contact mode AFM images (512×512 data points) of 5×5 $\mu m^2$, 2.5×2.5 $\mu m^2$, and 500×500 $nm^2$ are collected by keeping the working set point above 65% of the free oscillation amplitude. The scan rate for the acquisition of images was 0.6 Hz. Height profiles of ~ 100 flakes were processed by using JPK Data Processing software (JPK Instruments, Germany). The samples were prepared by drop-casting the exfoliated β-InSe flake dispersions onto mica sheets (G250-1, Agar Scientific Ltd., Essex, U.K.) and subsequently drying them under vacuum overnight.

Raman spectroscopy measurements were carried out using a Renishaw micro-Raman InVia with a 100× objective, an excitation wavelength of 514.5 nm and an incident power of 1 mW. For each sample, 20 spectra are collected. The samples were prepared by drop casting exfoliated β-InSe flake dispersions on Si/$SiO_2$ substrates and drying them under vacuum. The spectra were fitted with Lorentzian functions.

The crystal structure was characterized by XRD using a PANalytical Empyrean with Cu $K_\alpha$ radiation. The samples for XRD were prepared by drop-casting exfoliated β-InSe flake dispersions onto Si substrates and drying them under vacuum.

X-ray photoelectron spectroscopy characterization was carried out on a Kratos Axis UltraDLD spectrometer, using a monochromatic Al Kα source (15 kV, 20 mA). The spectra were acquired over an area of 300 μm x 700 μm. Wide scans were collected with a constant pass energy of 160 eV and an energy step of 1 eV. High-resolution spectra were acquired at a constant pass energy of 10 eV and an energy step of 0.1 eV. The binding energy scale was calibrated to the C 1s peak at 284.8 eV. The spectra were analysed using CasaXPS software (version 2.3.17). The fitting of the spectra was performed by using a Shirley background and Voigt profiles. The samples were prepared by drop-casting exfoliated β-InSe flake dispersions onto Si substrates and drying them under vacuum.

*Electrode fabrication*

Hybrid heterostructures were fabricated by the vacuum filtration of SWCNT (> 90% purity, Cheap Tubes) dispersions in 1-Methyl-2-pyrrolidinone (Reagent Grade, 99%, Sigma Aldrich) and exfoliated β-InSe flake dispersions on nylon membranes (Whatman® membrane filters nylon, pore size 0.2 μm). The mass fraction for both materials was kept constant at 1.2 mg cm$^{-2}$. The SWCNT dispersion was prepared by dispersing SWCNTs in NMP at a concentration of 0.4 g L$^{-1}$ using ultra-sonication based de-bundling.[114] The dispersion was sonicated using a horn probe sonic tip (Vibra-cell 75185, Sonics) with a vibration amplitude set to 45% and a sonication time of 30 min. The sonic tip was pulsed at a rate of 5 s on and 2 s off to avoid damage to the processor and to reduce any solvent heating. An ice bath was used during sonication in order to minimize heating effects.

*Electrode characterization*

High resolution scanning electron microscopy analysis was performed with a field-emission scanning electron microscope FE-SEM (Jeol JSM-7500 FA) in a high vacuum, with an acceleration voltage of 15 kV. Images were collected acquiring the backscattered electrons. A carbon coating of 10 nm on the sample is required to prevent charging effects.

Electrochemical measurements on the as-prepared hybrid heterostructures were carried out at RT in a flat-bottomed fused silica cell under a three-electrode configuration using a CompactStat potentiostat/galvanostat station (Ivium), controlled via IviumSoft. A Pt-wire was used as the counter electrode and saturated KCl Ag/AgCl was used as the reference electrode. Measurements were carried out in 0.5 M H$_2$SO$_4$ (99.999% purity, Sigma Aldrich) or 1 M KOH (≥ 85% purity, ACS reagent, pellets, Sigma Aldrich) aqueous electrolytes (pH = 1 and pH = 14, respectively). The potential difference between the working electrode and the Ag/AgCl reference electrode was converted to the reversible hydrogen electrode (RHE) scale via the Nernst equation: $E_{RHE} = E_{Ag/AgCl} + 0.059pH + E^0_{Ag/AgCl}$, in which $E_{RHE}$ is the converted potential versus RHE, $E_{Ag/AgCl}$ is the measured potential versus the Ag/AgCl reference electrode, and $E^0_{Ag/AgCl}$ is the standard potential of Ag/AgCl at 25 °C (0.1976 V). Linear sweep voltammetry curves were acquired at a scan rate of 5 mV/s and *iR*-corrected, in which *i* is the current and the *R* is the series resistance that arises from the substrate and electrolyte resistance. *R* is measured by electrochemical impedance spectroscopy (EIS) at an open circuit potential and frequency of $10^4$ Hz.

The linear portions of the Tafel plots fit the Tafel equation $\eta = m \log(j) + A$, in which $\eta$ is the overpotential with respect to RHE, *j* is the current density, *m* is the Tafel slope and *A* is the intercept of the linear regression. The *j₀* was calculated from the Tafel equation by setting $\eta = 0$.

*Theoretical methods*

The atomic structure and energetics of water splitting on InSe monolayers were studied by DFT using the QUANTUM-ESPRESSO code[115] and the GGA–PBE + van der Waals approximation, which are feasible for the description of the adsorption of molecules on surfaces.[116,117] Energy cut-offs of 25 Ry and 400 Ry were used for the plane-wave expansion of the wave functions and the charge density, respectively, and the 4×4×1 Monkhorst-Pack *k*-point grid for the Brillouin sampling.[118] Optimizations of both atomic positions and lattice vectors were performed for all calculations.

**Supporting Information**

Supporting Information is available from the Wiley Online Library or from the author.

**Acknowledgements**

#Elisa Petroni and Emanuele Lago contributed equally to this work.

The authors thank the European Commission's Horizon 2020 research and innovation programme under grant agreement No. 696656 – GrapheneCore1 for financial support, the Materials Characterization and Electron Microscopy Facilities at the Fondazione Istituto Italiano di Tecnologia, in particular Mr. Simone Lauciello for his help with HR-SEM measurements.

# Supporting Information

**Liquid-phase exfoliated indium-selenide flakes and their application in hydrogen evolution reaction**

*Elisa Petroni,[#] Emanuele Lago,[#] Sebastiano Bellani, Danil W. Boukhvalov, Antonio Politano, Bekir Gürbulak, Songül Duman, Mirko Prato, Silvia Gentiluomo, Reinier Oropeza-Nuñez, Jaya-Kumar Panda, Peter S. Toth, Antonio Esau Del Rio Castillo, Vittorio Pellegrini, and Francesco Bonaccorso\**

**Supplementary Text**

*Growth of InSe single crystals*

InSe crystals were grown using the Bridgman-Stockbarger method. The melting point of the InSe compound, namely 933 ± 5 K, was determined from the phase diagram. In and Se starting elements of a high purity grade (not less than 99.999%) were sealed in quartz ampoule, which was annealed at 1220 K under a vacuum of $10^{-6}$ mbar for 15 hours in a furnace. The temperature of the quartz ampoule was decreased to RT in 24 hours. The crucible was then suspended in the middle of the vertical furnace. The temperature of the furnace was increased to 1220 K and maintained at this temperature for 40 hours, then decreased to 1023 K. The temperature was kept at 1023 K for 15 hours. The temperature of the lower zone of furnace was reduced to 520 K at a rate of 1.54 K h$^{-1}$. Both the furnace zones were cooled to 520 K in 75 hours. The solidified ingot was cooled to RT in 50 hours. The grown InSe single-crystals samples were 10 mm in diameter and about 60 mm in length.

**Supplementary Figures**

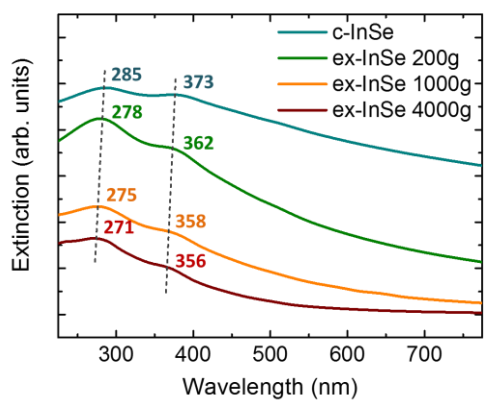

**Figure S1** Extinction spectra of c-InSe (cyan curve), ex-InSe 200*g* (green curve), ex-InSe 1000*g* (orange curve) and ex-InSe 4000*g* (red curve). The dashed lines highlight the blue shift in position of direct transition peaks from bulk to the exfoliated InSe samples.

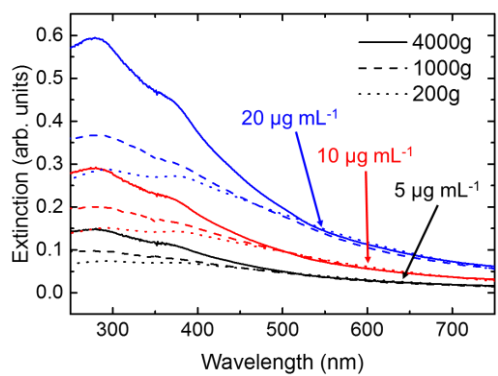

**Figure S2** Extinction spectra of ex-InSe samples at different centrifugation speeds (200*g*, dotted curves, 1000*g*, dashed curves, and 4000*g*, continue curves) and different concentrations (5μg mL$^{-1}$, black curves, 10μg mL$^{-1}$, red curves, 20μg mL$^{-1}$, blue curves). It is worth noting that the curves of the samples with the same concentration overlap below 550 nm.

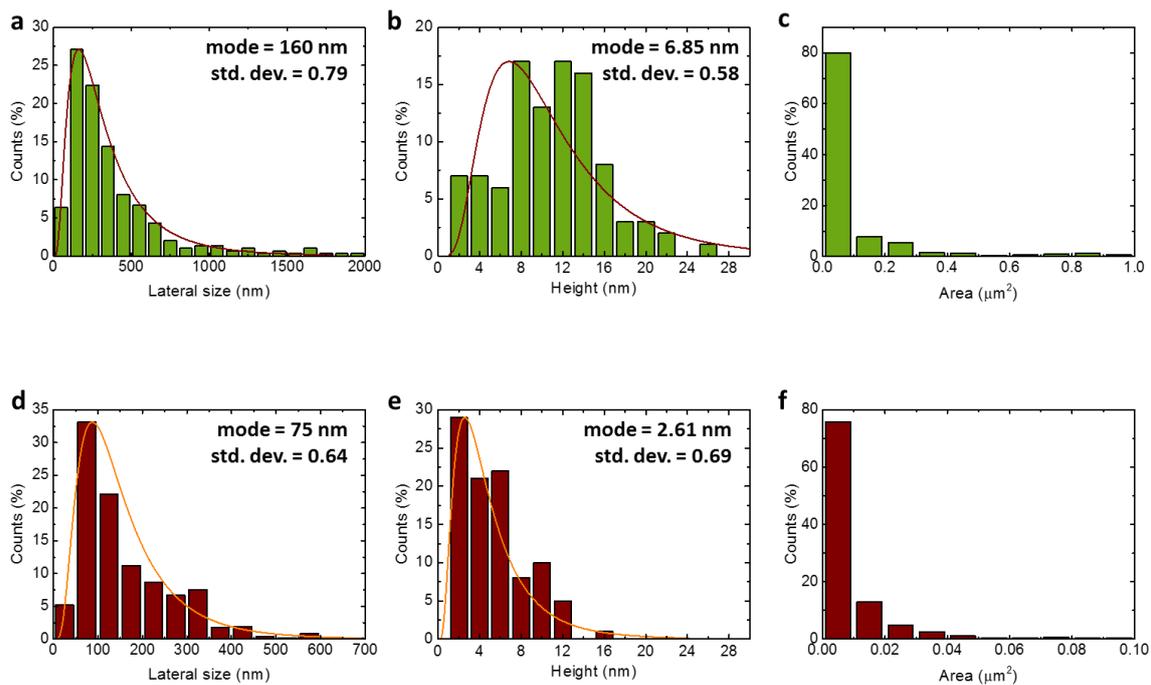

**Figure S3** Statistical morphological analysis of ex-InSe samples. **a** Lateral size, **b** height and **c** area of ex-InSe 200*g*; **d** Lateral size, **e** height and **f** area of ex-InSe 4000*g*.

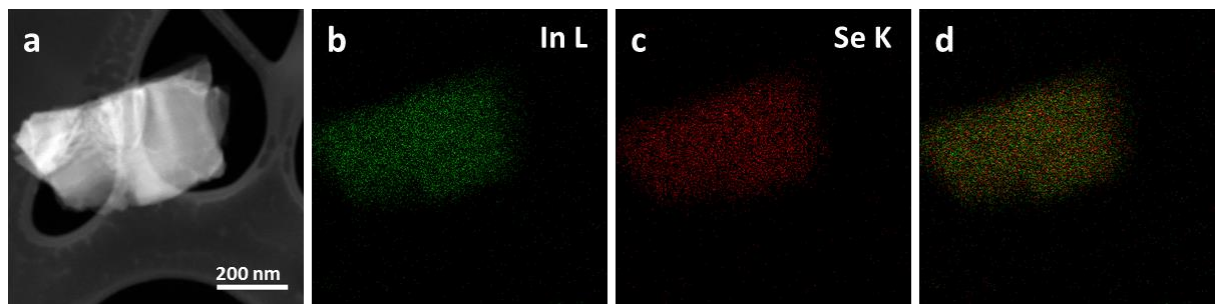

**Figure S4** STEM – EDS analysis of an ex-InSe 1000*g* flake. **a** STEM image of an ex-InSe flake; **b** In L map; **c** Se K map; **d** overlap between the In L and Se K maps.

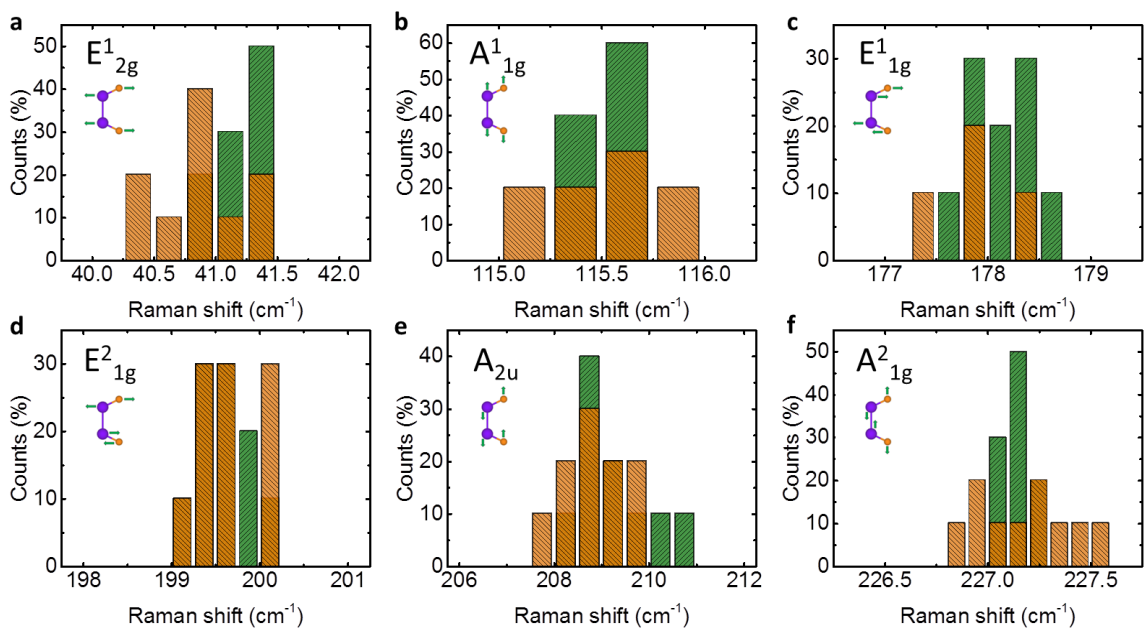

**Figure S5** Raman statistical analysis of c-InSe (green bars) and ex-InSe 1000g (orange bars) samples. **a** pos($E^1_{2g}$); **b** pos($A^1_{1g}$); **c** pos($E^1_{1g}$); **d** pos($E^2_{1g}$); **e** pos($A_{2u}$); **f** pos($A^2_{1g}$). No energy shifts between the bulk (c-InSe) and exfoliated (ex-InSe 1000*g*) samples are observed. Detail of Raman active atomic vibrations of InSe in insets.

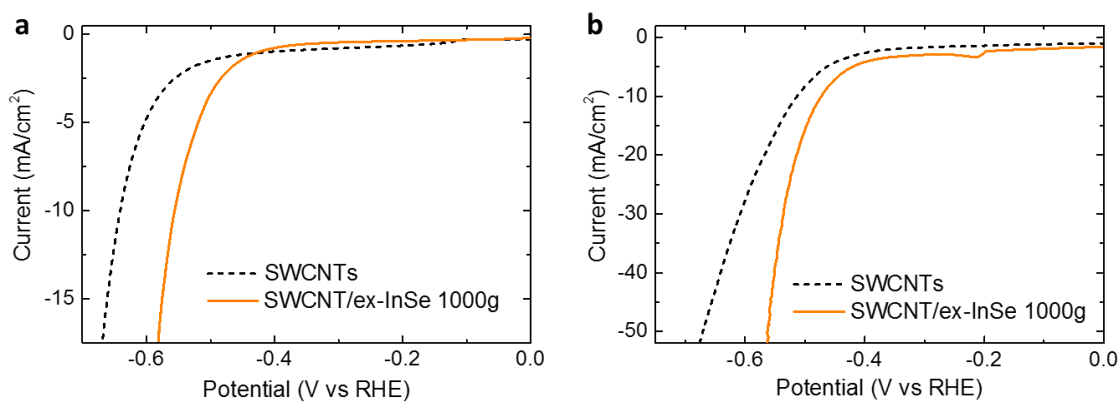

**Figure S6** Comparison of HER activity between SWCNT-based membrane and SWCNT/ex-InSe 1000*g* hybrid heterostructure. **a** linear sweep voltammetries acquired at pH = 1; **b** linear sweep voltammetries acquired at pH = 14.

**Table ST1** Calculated $j_0$ values for c-InSe and ex-InSe samples in both acid and alkaline solutions.

| electrolyte | $j_0$ (µA cm$^{-2}$) | | | |
|---|---|---|---|---|
| | c-InSe | ex-InSe 200*g* | ex-InSe 1000*g* | ex-InSe 4000*g* |
| Acid (pH = 1) | 0.216 | 0.177 | 0.227 | 0.419 |
| Alkaline (pH = 14) | 9.379 | 0.448 | 0.602 | 5.841 |

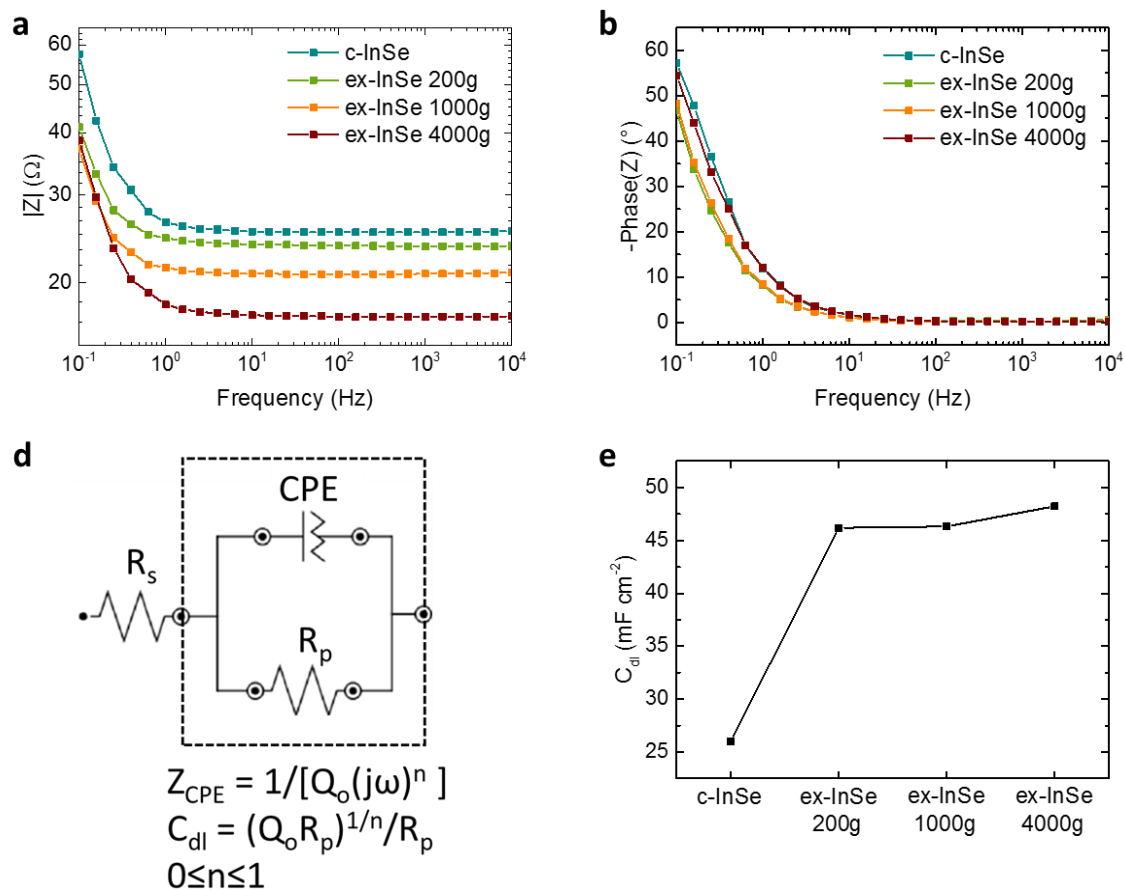

**Figure S7** Electrochemical impedance bode plots for **a** impedance magnitude (|Z|) and **b** the negative impedance phase (-phase(Z) ) vs. frequency for SWCNT/c-InSe and SWCNT/ex-InSe (200*g*, 1000*g* and 4000*g* samples) in 0.5 M H$_2$SO$_4$; **c** Equivalent circuit used for extrapolating the double layer capacitance at the InSe/electrolyte interface ($C_{dl}$). $R_s$ is the series resistance for the electrode and the electrolyte, while the parallel between the constant phase element (CPE) and the resistance $R_p$ is used for representing $C_{dl}$. The equation for the impedance of CPE ($Z_{CPE}$) and the $C_{dl}$ are also reported. $Q_0$ and n ($0 \leq n \leq 1$) are frequency

independent parameters; **d** Plot for $C_{dl}$ vs. c-InSe and ex-InSe (200*g*, 1000*g* and 4000*g* samples).